\documentclass[11pt,fleqn]{article}
\usepackage{amsmath,amssymb,amsthm,cite,enumerate,graphicx}

\allowdisplaybreaks

\newcommand{\setSkips}{\vspace*{-1.0mm plus 0mm minus 0.2mm}\parskip=0mm\parsep=0mm\itemsep=0.1mm\topsep=-0.1mm}
\newcommand{\const}{\mathop{\rm const}\nolimits}
\newcommand{\dt}{\,{\rm d}t}
\newcommand{\noprint}[1]{}

\newtheorem{proposition}{Proposition}
{\theoremstyle{definition}

\newtheorem{remark}{Remark}

}

\begin{document}

{\parindent=0mm

{\LARGE\bf Equivalence groupoid \\ of generalized potential Burgers equations\par}

\vspace{3.2mm}

{\large Oleksandr A. Pocheketa}

\vspace{1.4mm}

\emph{Institute of Mathematics of NAS of Ukraine,~3 Tereshchenkivska Str., 01601 Kyiv, Ukraine}

\vspace{0.2mm}

Email: pocheketa@yandex.ua

}

{\vspace{5mm}\par\noindent\hspace*{8mm}\parbox{144mm}{\small
We find the equivalence groupoid of a~class
of $(1+1)$-dimensional second-order evolution equations, which are called generalized potential Burgers equations.
This class is related via potentialization with two classes of variable-coefficient generalized Burgers equations.
Its equivalence groupoid is of complicated structure
and is described via partitioning the entire class into three normalized subclasses
such that there are no point transformations between equations from different subclasses.
For each of these subclasses we construct its equivalence group of an appropriate kind.
}\par\vspace{2mm}}


\section{Introduction}
The classical Burgers equation $u_t+uu_x-u_{xx}=0$ was generalized in various ways including variable-coefficient generalizations.
Some of them were studied in the framework
of symmetry analysis \cite{katk1965a,doyl1990a,king1991c,boyk1997a,wafo2004d,soph2004a,poch2012a,poch2013d,vane2015a}.

In this paper we study admissible transformations of the class of $(1+1)$-dimensional second-order evolution equations of the general form
\begin{gather}\label{PotClass}
v_{t}+v_{x}^2+f(t,x)v_{xx}=0, \quad f\ne0.
\end{gather}
We denote an equation of the form \eqref{PotClass} with a~fixed value of the arbitrary element~$f$ by~$\mathcal{P}_f$,
and call it a~{\em generalized potential Burgers equation} for two reasons.
The first one is that~$\mathcal{P}_f$ is a~potential equation for the equation
\begin{gather}\label{ClassDiffCoef}
u_t+2uu_x+\big(f(t,x)u_x\big)_x=0
\end{gather}
with the same $f\ne0$.
For any value of the arbitrary element~$f$, the space of conservation laws of the corresponding equation $\mathcal{C}_f$
of the form \eqref{ClassDiffCoef}
is one-dimensional and is spanned by the obvious conservation law with the characteristic $1$.
Hence the associated potential system has the form $v_x=u$, $v_t=-u^2-fu_x$, which leads to the equation~$\mathcal{P}_f$
for the potential~$v$.

The second reason is that an~equation~$\mathcal{P}_{\hat f}$ in hatted variables,
$\hat v_{\hat t}+\hat v_{\hat x}^2+\hat f(\hat t,\hat x)\hat v_{\hat x\hat x}=0$, with $\smash{\hat f_{\hat x\hat x\hat x}=0}$
is similar with respect to a~point transformation
to the potential equation for an equation~$\mathcal{L}_f$ with $f_{xxx}=0$ from the class
\begin{gather}\label{GBE}
u_t+uu_x+f(t,x)u_{xx}=0, \quad f\ne0,
\end{gather}
which is called the class of {\em generalized Burgers equations} \cite{poch2012a}.
The equation~$\mathcal{L}_f$ has nonzero conservation laws if and only if~$f_{xxx}=0$, and in this case the space of conservation laws
is one-dimensional. It is spanned by the conservation law with the characteristic $\lambda=e^{\int\!f_{xx}\dt}$,
which depends only on~$t$, see~\cite{poch2016a}.
Here and in what follows an integral with respect to~$t$ means a~fixed antiderivative.
The potential system constructed for~$\mathcal{L}_f$ with respect to the
canonical conserved current of the above conservation law is
\begin{gather*}
v_x=\lambda u,
\quad
v_t=-\lambda\left(\frac12u^2+fu_x-f_xu\right)\!,
\end{gather*}
and the corresponding potential equation is
\begin{gather*}
v_t+\frac1{2\lambda}v_x^2+fv_{xx}-f_xv_x=0.
\end{gather*}
Under the assumption~$f_{xxx}=0$, the last equation is similar,
with respect to the transformation
\begin{gather*}
\hat t=\frac12\int\lambda \dt,
\quad
\hat x=\lambda x+\int f^1\lambda \dt,
\quad
\hat v=v,
\end{gather*}
to the equation~$\mathcal{P}_{\hat f}$, where
$\hat f(\hat t,\hat x)=2\lambda f(t,x)$ and $f^1=f_x-xf_{xx}$ is a~function of~$t$ only.

Note that the intersection of the classes \eqref{ClassDiffCoef} and~\eqref{GBE} is their subclass
consisting of the equations with $f=f(t)$.
Traditionally, such equations
have received more attention due to their appearance in physical models
(see~\cite{crig1979a,scot1981a,hamm1989a,hamm1989b,sion1994a,sach1994a} and also~\cite[Chapter~4]{leib1974a}).
We point out that it is the first class of differential equations that was considered in order to determine
its equivalence groupoid~\cite{king1991c}.
The symmetry analysis of this class was carried out in~\cite{doyl1990a,wafo2004d}
and was enhanced and completed in~\cite{poch2014a}.

In general, the further potentialization of equations from the class~\eqref{PotClass} is not possible
since any equation~$\mathcal{P}_f$ with $f\ne\const$ has no nonzero conservation laws.
At the same time, equations with constant values of the arbitrary element~$f$ are singular in the class~\eqref{PotClass}
from the point of view of conservation laws and transformational properties,
and they constitute the orbit of the potential Burgers equation~$\mathcal{P}_{-1}$.
It is well known that for each equation~$\mathcal{P}_f$ with $f=\const$,
its maximal invariance algebra is infinite-dimensional~\cite[Example~2.42]{olve1993b}.
The space of conservation laws of~$\mathcal{P}_f$ is also infinite-dimensional and is spanned by conservation laws with the characteristics
$\mu=\psi\exp({\frac{v}{2f}})$, where the parameter-function $\psi=\psi(t,x)$
runs through the solution set of the linear heat equation $\psi_t=f\psi_{xx}$~\cite{popo2005b}.
This is explained by the fact that the equation~$\mathcal{P}_f$ with $f=\const$
can be linearized to the ``backward'' heat equation $\tilde v_t+f\tilde v_{xx}=0$ by means of the variable change $\tilde v=e^{v/f}$.
Moreover, this essentially complicates the structure of the equivalence groupoid of the entire class~\eqref{PotClass},
cf.~Proposition~\ref{Prop_PotClassConst_G} below, and makes the study challenging.

Admissible transformations considered with the natural operation of transformation compo\-si\-tion
constitute the so-called equivalence groupoid of the class~\eqref{PotClass}.
The notion of equivalence groupoid was suggested in~\cite{popo2012a,bihl2012b}.
In order to describe the equivalence groupoid of the class~\eqref{PotClass}
we apply the technique based on partitioning the class~\eqref{PotClass}
into normalized (in the usual, or in the generalized, or in the generalized extended sense) subclasses, cf.\ \cite{popo2006b,popo2010a}.
Equations of different subclasses are not similar with respect to point transformations.
For each of the subclasses,
we construct the corresponding equivalence group of the same kind
(usual, generalized, or generalized extended)
as the kind of normalization of the subclass.
We also comprehensively compare the equivalence groupoids of the classes~\eqref{PotClass}, \eqref{ClassDiffCoef} and~\eqref{GBE}.

Definitions of admissible transformations, normalized and semi-normalized classes of dif\-fe\-ren\-tial equations,
various kinds of equivalence groups
and other concepts and techniques related to group classification problems
can be found in \cite{popo2010a,ivan2010a,vane2009a} and references therein.

\section{Equivalence groupoid}\label{Section_P_Propositions}

Transformational properties in the class~\eqref{PotClass} are exhaustively described by the following chain of assertions.
These assertions result from a~long calculation procedure, and thus it is convenient to place their proofs in a~separate section.

\begin{proposition}\label{Prop_PotClass_G}
The usual equivalence group~$G^\sim_{\mathrm{pot}}$
of the class~\eqref{PotClass} consists of the transformations
\begin{gather}\label{PotClass_G}
\tilde t=\alpha t+\beta,
\quad
\tilde{x}=\kappa(x+\mu_1t+\mu_0),
\quad
\tilde{v}=\frac{\kappa^2}{\alpha}\left(v+\frac{\mu_1}{2}x+\frac{\mu_1^2}{4}t+\nu\right),
\quad
\tilde f=\frac{\kappa^2}{\alpha}f,
\end{gather}
where $\alpha$, $\beta$, $\kappa$, $\mu_1$, $\mu_0$, $\nu$ are arbitrary constants with $\alpha\ne0$ and $\kappa\ne0$.
\end{proposition}

We point out that the generalized extended equivalence group of this class coincides with its usual equivalence group.

\begin{proposition}\label{Prop_PotClassConst_G}
The generalized equivalence group of the subclass~$\mathcal{P}^3=\{\mathcal{P}_f \mid f=\const\}$
of the class~\eqref{PotClass} is formed by the transformations
\begin{gather}\label{PotClassConst_G}
\tilde{t}=\frac{\alpha t+\beta}{\gamma t+\delta},
\quad
\tilde{x}=\kappa\frac{x+\mu_1t+\mu_0}{\gamma t+\delta},
\nonumber
\\
\tilde{v}=\frac{\kappa^2f}{\alpha\delta-\beta\gamma}\ln\big|F^1\big(e^{v/f}+F^2\big)\big|,
\quad
\tilde{f}=\frac{\kappa^2}{\alpha\delta-\beta\gamma}f,
\end{gather}
where $\alpha$, $\beta$, $\gamma$, $\delta$, $\kappa$, $\mu_1$, $\mu_0$ are arbitrary constants with $\alpha\delta-\beta\gamma\ne0$ and $\kappa\ne0$;
the tuple  $(\alpha,\beta,\gamma,\delta,\kappa)$ is defined up to a~nonzero multiplier;
\begin{gather}\label{F1}
F^1=\begin{cases}
k\sqrt{|\gamma t+\delta|}\exp\left(-\dfrac{(\gamma x-\mu_1\delta+\mu_0\gamma)^2}
{4f\gamma(\gamma t+\delta)}\right), & \gamma\ne0,
\\[4mm]
k\exp\dfrac{2\mu_1x+\mu_1^2t}{4f}, & \gamma=0,
\end{cases}
\end{gather}
$k$ is a~nonzero constant, and $F^2$ is a~solution of the linear equation
\begin{gather}\label{F2}
F^2_t+fF^2_{xx}=0.
\end{gather}
The subclass $\mathcal{P}^3$ is normalized in the generalized sense.
\end{proposition}

The usual equivalence group of the subclass~$\mathcal{P}^3$
coincides with~$G^\sim_{\mathrm{pot}}$,
but its generalized equivalence group is wider.
Therefore, the subclass~$\mathcal{P}^3$ is not normalized in the usual sense.
Moreover, this also implies that in the subclass~$\mathcal{P}^3$ there exist admissible transformations
that are not induced by $G^\sim_{\mathrm{pot}}$, which is the generalized extended equivalence group of the class~\eqref{PotClass}.
This means that the class~\eqref{PotClass} is not normalized in any sense.

Proposition~\ref{Prop_PotClassConst_G} shows that for any constant $f\ne0$ the equation~$\mathcal{P}_f$ has
rather complicated point symmetry transformations that are not related to equivalence transformations of the class~\eqref{PotClass};
one may substitute $\kappa^2=\alpha\delta-\beta\gamma$ into~\eqref{PotClassConst_G}
to derive the form of symmetry transformations of~$\mathcal{P}_f$ and observe the above fact.
It is easy to check that any admissible transformation in the subclass~$\mathcal{P}^3$ is induced
by the composition of a~symmetry transformation of the corresponding initial equation
and a~transformation from~$\smash{G^\sim_{\mathrm{pot}}}$.
In other words, the subclass~$\mathcal{P}^3$ is semi-normalized in the usual sense.
This is a~consequence of that the subclass~$\mathcal{P}^3$ is the single orbit of any equation from~$\mathcal{P}^3$
with respect to its usual equivalence group~$G^\sim_{\mathrm{pot}}$, cf.\ the
proof of~\cite[Proposition~2]{boyk2014a}.

\begin{proposition}\label{Prop_PotClass_NonQuad_G}
The usual equivalence group of the subclass~$\mathcal{P}^1=\{\mathcal{P}_f \mid f_{xxx}\ne0\}$
of the class~\eqref{PotClass} coincides with the usual equivalence group~$G^\sim_{\mathrm{pot}}$
of the entire class~\eqref{PotClass}
and, thus, it is governed by~\eqref{PotClass_G}.
The subclass $\mathcal{P}^1$ is normalized in the usual sense.
\end{proposition}

Although the class~\eqref{PotClass} is not normalized, it can be partitioned
into three subclasses that are normalized (in certain senses),
namely~$\mathcal{P}^1$,~$\mathcal{P}^3$, and the complement of their union,~$\mathcal{P}^2=\overline{\mathcal{P}^1\cup\mathcal{P}^3}$.
The subclass $\overline{\mathcal{P}^1}$, which is singled out from the class~\eqref{PotClass},
by the constraint $f_{xxx}=0$ has complicated transformation properties and is not normalized.
At the same time,
subtracting the small subclass~$\mathcal{P}^3$ from $\overline{\mathcal{P}^1}$
we obtain the normalized subclass~$\mathcal{P}^2$.

\begin{proposition}\label{Prop_PotClass_Quad_G}
The generalized extended equivalence group of the subclass~$\mathcal{P}^2=\{\mathcal{P}_f \mid f_{xxx}=0$, $f\ne\const\}$
of the class~\eqref{PotClass} is constituted by the transformations of the form
\begin{gather}
\tilde t=\frac{1}{c_0}\!\int\!(X^1)^2\dt+c_5,
\quad
\tilde x=X^1x+c_1\!\int\! (X^1)^2\left(\int\!\frac{f^1}{\lambda}\dt+c_3\right)\!\dt+c_4,
\nonumber
\\
\tilde v=c_0\left(v-\frac{c_1X^1}{4\lambda}x^2+X^2x+\int\!\big(X^2\big)^2\dt
+\frac{c_1}{2}\!\int\!\frac{f^0}{\lambda}X^1\dt\right)\!+c_5,
\quad
\tilde f=c_0f,
\label{PotClass_Quad_G}
\end{gather}
where $f=f^2(t)x^2+f^1(t)x+f^0(t)$,
\begin{gather*}
X^1(t):=\left(\!c_1\!\int\!\frac{\dt}{\lambda}+c_2\!\right)^{-1},
\quad
X^2(t):=\frac{c_1X^1\!}{2}\left(\int\!\frac{f^1}{\lambda}\dt+c_3\!\right),
\quad
\lambda(t):=e^{2\int f^2\dt},
\end{gather*}
and  $c_0,\ldots,c_5$ are arbitrary constants with $c_0\ne0$ and $(c_1,c_2)\ne(0,0)$.

The subclass $\mathcal{P}^2$ is normalized in the generalized extended sense.
\end{proposition}

The usual equivalence group of the subclass~$\mathcal{P}^2$ also
coincides with~$G^\sim_{\mathrm{pot}}$, but it does not generate all admissible transformations in this subclass.

\begin{remark}
Two approaches can be used here when we represent the generalized extended equivalence group of~$\mathcal{P}^2$.
On the one hand, if we consider the function~$f$ as the single arbitrary element of the subclass~$\mathcal{P}^1$,
then we should explicitly express the coefficients $f^2$, $f^1$ and $f^0$ of the polynomial $f=f^2(t)x^2+f^1(t)x+f^0(t)$
in terms of~$f$ and its derivatives,
\begin{gather*}
f^2=\frac12 f_{xx},
\quad
f^1=f_x-xf_{xx},
\quad
f^0=f+\frac{x^2}{2}f_{xx}-xf_x ,
\end{gather*}
and substitute these expressions into \eqref{PotClass_Quad_G}.
On the other hand, we may assume that the three functions $f^2$, $f^1$ and~$f^0$ (or, e.g., $\lambda$, $f^1$ and $f^0$)
are the arbitrary elements of the class~\eqref{PotClass}
and perform a~reparametrization of this class.
In particular, the equation $\tilde f=c_0f$ should be considered
as $\tilde f^2(\tilde t)\tilde x^2+\tilde f^1(\tilde t)\tilde x+\tilde f^0(\tilde t)=c_0\big(f^2(t)x^2+f^1(t)x+f^0(t)\big)$,
and then it should be split with respect to~$x$ in order to derive
explicit expressions for~$\tilde f^2$,~$\tilde f^1$ and~$\tilde f^0$ in terms of~$f^2$,~$f^1$ and~$f^0$.
In this paper, we prefer to avoid any reparametrization, and thus
the notation of $\lambda$ serves only for minimizing the size of expressions.
\end{remark}

\begin{remark}
Note that
in the case when $f^2=0$ (i.e. $f=f^1(t)x+f^0(t)$), we have $\lambda=1$. Hence
the expressions \eqref{PotClass_Quad_G} after redenoting the constants take the form
\begin{gather*}
\tilde t=\frac{\alpha t+\beta}{\gamma t+\delta},
\quad
\tilde x=\frac{\kappa(x+\nu\Delta^{-1}(\alpha t+\beta))}{\gamma t+\delta}+\gamma\kappa\!\int\!\frac{\!\int\! f^1\dt}{(\gamma t+\delta)^2}\dt+c_4,
\\[.25ex]
\tilde v=\frac{\kappa^2}{\Delta}\left(v
-\frac{\gamma x^2 + 2\big[\gamma\int\! f^1\dt+\nu\big]x}{4(\gamma t+\delta)}
 + \frac{1}{4}\!\int\!\left[\frac{\gamma\int\! f^1\dt+\nu}{(\gamma t+\delta)}\right]^2\!\!\dt
 + \frac{\gamma}{2}\!\int\!\frac{f^0\dt}{\gamma t+\delta} \right)+c_5,
\\
\tilde f=\frac{\kappa^2}{\Delta}f,
\end{gather*}
where $\Delta=\alpha\delta-\beta\gamma \ne 0$ and the tuple $(\alpha,\beta,\gamma,\delta,\kappa)$ is defined up to a~nonzero multiplier.
Further simplification is possible if $f^1=0$.
\end{remark}

\begin{proposition}\label{Prop_PotClass_subclasses}
There are no point transformations between any two equations from different subclasses~$\mathcal{P}^1$,
$\mathcal{P}^2$ or $\mathcal{P}^3$ of the class~\eqref{PotClass}.
\end{proposition}

Proposition~\ref{Prop_PotClass_subclasses} completes
the description of the equivalence groupoid of the class~\eqref{PotClass}, cf.~\cite[Section 3.4]{popo2010a}.

Recall that the equivalence group of a~subclass of a~class of differential equations
is called a~{\em conditional equivalence group} of the entire class~\cite{popo2010a,popo2006b}.
Only maximal conditional equivalence groups are essential~\cite[Definition~7]{popo2010a}.
Therefore, Propositions~\ref{Prop_PotClassConst_G} and \ref{Prop_PotClass_Quad_G} mean that
the class \eqref{PotClass} admits nontrivial conditional equivalence groups (in the generalized or the generalized extended sense).
Taking into account the results of \cite[Section 3.4]{popo2010a} and Proposition~\ref{Prop_PotClass_subclasses}
we can also claim that the maximal conditional equivalence groups of the class~\eqref{PotClass}
are exhausted by the above conditional equivalence groups
and the equivalence group of the entire class~\eqref{PotClass}.

\section{Proofs}

Given an equation~$\mathcal{P}_f$ of the form~\eqref{PotClass} and another (``tilded'') equation from the same class,
$\tilde v_{\tilde t}+\tilde v_{\tilde x}^2+\tilde f\tilde v_{\tilde x\tilde x}=0$,
we consider point transformations between these equations.
In view of \cite[Lem\-ma~1]{ivan2010a} these transformations are of the form $\tilde t=T(t)$, $\tilde x=X(t,x)$, $\tilde v=V(t,x,v)$,
where $T_tX_xV_v\ne0$.
Expressing in the ``tilded'' equation all the tilded entities in terms of the untilded ones,
\begin{gather*}
\tilde v_{\tilde t}=\frac{1}{T_t}\left(V_t+V_v v_t-\frac{V_x+V_v v_x X_t}{X_x}\right),
\\
\tilde v_{\tilde x}=\frac{V_x+V_v v_x}{X_x},
\quad
\tilde v_{\tilde x \tilde x}=\frac{V_{xx}+2V_{vx}v_x+V_{vv}v_x^2+V_v v_{xx}}{X_x^2} - \frac{X_{xx}V_x+X_{xx}V_v v_x}{X_x^3},
\end{gather*}

\noindent
and  confining it to the manifold defined by the equation~$\mathcal{P}_f$
via substituting $v_t=-v_{x}^2-fv_{xx}$, we split the result with respect to derivatives of~$v$.
This gives $\tilde f={X_x^2}f/T_t$ and the classifying equations
\begin{gather}
V_v^2-\frac{X_x^2}{T_t}V_v+f\frac{X_x^2}{T_t}V_{vv}=0,
\label{GagedPEdeq2}
\\
2V_xV_v-\frac{X_tX_x}{T_t}V_v+2f\frac{X_x^2}{T_t}V_{xv}-f\frac{X_xX_{xx}}{T_t}V_v=0,
\label{GagedPEdeq1}
\\
V_t\frac{X_x^2}{T_t}-\frac{X_tX_x}{T_t}V_x+V_x^2+f\frac{X_x^2}{T_t}V_{xx}-f\frac{X_xX_{xx}}{T_t}V_x=0.
\label{GagedPEdeq0}
\end{gather}

In order to find the usual equivalence group for the class~\eqref{PotClass},
we vary~$f$ and split~\eqref{GagedPEdeq2}--\eqref{GagedPEdeq0} with respect to~it.
Solving the obtained system of determining equations proves Proposition~\ref{Prop_PotClass_G}.

However, we are interested in further investigation of transformational properties of the class~\eqref{PotClass},
so we continue to analyze the equations~\eqref{GagedPEdeq2}--\eqref{GagedPEdeq0}.
Integrating~\eqref{GagedPEdeq2} we get
$V=\tilde f\ln\left| F^1(t,x)(e^{v/f}+F^2(t,x)) \right|$,
where $F^1$ and $F^2$ are smooth functions of their arguments, and $F^1(t,x)\ne0$ since $V_v\ne0$.
There are two essentially different cases for admissible transformations of the class~\eqref{PotClass}
depending on whether $F^2(t,x)=0$.

\textbf{Case 1.}
Under the assumption $F^2\ne0$, the functions $1$, $e^{v/f}$, $ve^{v/f}$, $\ln|F^1(e^{v/f}+F^2)|$ and~$\ln|F^1(e^{v/f}+F^2)|e^{v/f}$
are linearly independent as functions of~$v$ over the ring of smooth functions of $(t,x)$.
Substituting the obtained expression for $V$ into the equation \eqref{GagedPEdeq1} and splitting it with respect to the above functions
we obtain $f_x=0$ and $X_{xx}=0$, so we can represent $X$ as $X=X^1(t)x+X^0(t)$, $X^1\ne0$.
In the same way, from the determining equation~\eqref{GagedPEdeq0} we have {\samepage
\begin{gather*}
f_t=0,
\quad
2X^1_t=\frac{T_{tt}}{T_t}X^1,
\quad
2fX^1F^1_x=(X^1_tx+X^0_t)F^1,
\\
X^1(F^1_t+fF^1_{xx})=(X^1_tx+X^0_t)F^1_x,
\quad
X^1\big((F^1F^2)_t+f(F^1F^2)_{xx}\big)=(X^1_tx+X^0_t)(F^1F^2)_x,
\end{gather*}
where the last three equations are already simplified in view of the condition $f=\const$. }

Integrating the second and the third equations we obtain
\begin{gather*}
T_t=c_0(X^1)^2,
\quad
F^1=K(t)\exp\left({\frac{X^1_tx^2+2X^0_tx}{4fX^1}}\right),
\end{gather*}
where $c_0$ is a~nonzero constant and $K$ is a~nonzero smooth function of~$t$ to be defined.
Thus, the fourth equation takes the form
\begin{gather}
\frac{K_t}{K}=-\frac{X^1}{4f}\left(\frac{X^1_t}{(X^1)^2}\right)_t x^2
-\frac{X^1}{2f}\left(\frac{X^0_t}{(X^1)^2}\right)_t x+\frac1{4f}\left(\frac{X^0_t}{X^1}\right)^2-\frac{X^1_t}{2X^1}.
\label{DetEq4}
\end{gather}
Since the ratio ${K_t}/{K}$ in the left-hand side
is a~function of~$t$, we conclude that the coefficients of~$x$ and of $x^2$ in~\eqref{DetEq4} necessarily vanish.
As a~result, we derive the forms of $X^1$, $X^0$ and $T$,
\begin{gather*}
X^1(t)=\frac{\kappa}{\gamma t+\delta},
\quad
X^0(t)=\kappa\frac{\mu_1t+\mu_0}{\gamma t+\delta},
\quad
T(t)=\frac{\alpha t+\beta}{\gamma t+\delta},
\end{gather*}
where $\alpha$, $\beta$, $\gamma$, $\delta$, $\mu_0$, $\mu_1$ and $\kappa$ are arbitrary constants
satisfying conditions of Proposition~\ref{Prop_PotClassConst_G}.
Then equation \eqref{DetEq4} is rewritten as
\begin{gather*}
\frac{K_t}{K}=\frac{(\mu_1\delta-\mu_0\gamma)^2}{4f(\gamma t+\delta)^2}+\frac{\gamma}{2(\gamma t+\delta)},
\end{gather*}
which can be easily integrated. The general solution is
\begin{gather*}
K=\begin{cases}
k\sqrt{|\gamma t+\delta|}\exp\left(-\dfrac{(\mu_1\delta-\mu_0\gamma)^2}{4f\gamma(\gamma t+\delta)}\right), & \gamma\ne0,
\\
k\exp\left(\dfrac{\mu_1^2}{4f}t\right), & \gamma=0,
\end{cases}
\end{gather*}
where the constant $k$ is nonzero.
Hence $F^1$ is of the form \eqref{F1} and $F^2$ is a~solution of \eqref{F2}.
Thus, Proposition~\ref{Prop_PotClassConst_G} holds.

\textbf{Case 2.}
If $F^2=0$, then $V$ can be represented in the form $V=\dfrac{X_x^2}{T_t}v+V^0$,
where $V^0$ is a~smooth function of $(t,x)$, which can be expressed in terms of $F^1$ (the precise expression is not essential).
Equation~\eqref{GagedPEdeq2} becomes an identity.
Splitting equations~\eqref{GagedPEdeq1}--\eqref{GagedPEdeq0}
with respect to~$v$ we obtain
\begin{gather}
X_{xx}=0,
\quad
\text{hence}
\quad
X=X^1(t)x+X^0(t),
\nonumber
\\
2X^1_tT_t-X^1T_{tt}=0,
\quad
\text{hence}
\quad
(X^1)^2=c_0T_t,
\quad
c_0\ne0,
\nonumber
\\
V^0_x=\frac{X^1}{2T_t}(X^1_t x+X^0_t),
\quad
\text{hence}
\quad
V^0=\frac{c_0X^1_t}{4X^1}x^2+\frac{c_0X^0_t}{2X^1}x+V^{00}(t),
\nonumber
\\
\big(X^1X^1_{tt}-2(X^1_t)^2\big)x^2+2\big(X^1X^0_{tt}-2X^1_tX^0_t\big)x+2fX^1X^1_t+4T_tV^{00}_t-(X^0_t)^2=0.
\label{GagedPEdeqLinV}
\end{gather}

If $f_{xxx}\ne0$, then the functions $1$,~$x$, $x^2$ and~$f$ are linearly independent as functions of~$x$
over the ring of smooth functions of~$t$.
Thereby the last equation of system~\eqref{GagedPEdeqLinV} admits splitting with respect to these functions.
Integrating the complete system of its consequences
we get $X^1=\kappa$, $X^0=\kappa(\mu_1t+\mu_0)$,
$T(t)=\alpha t+\beta$, $V=\frac{\kappa^2}{\alpha}\big(v+\frac{\mu_1}{2}x+\frac{\mu_1^2}{4}t+\nu\big)$ and
$\tilde f=\frac{\kappa^2}{\alpha}f$,
where $\alpha:=\frac{\kappa^2}{c_0}$, $\beta$, $\mu_1$, $\mu_0$, $\nu$ and $\kappa$ are arbitrary constants with $\alpha\kappa\ne0$.
As a~result, Proposition~\ref{Prop_PotClass_NonQuad_G} is derived.

In the case $f_{xxx}=0$ we represent~$f$ as $f=f^2(t)x^2+f^1(t)x+f^0(t)$, where $f^2$, $f^1$ and $f^0$
are smooth functions of~$t$, and the last equation of \eqref{GagedPEdeqLinV} after splitting with respect to~$x$ gives
\begin{gather*}
2f^2+\frac{X^1_{tt}}{X^1_t}-2\frac{X^1_t}{X^1}=0,
\quad
\left(\frac{X^0_{tt}}{X^1_t}\right)_t=-\frac{f^1X^1}{(X^1)^2},
\quad
4(X^1)^2=c_0(X^0)^2-2c_0f^0X^1{X^1_t}.
\end{gather*}
These equations can be easily integrated by quadratures, and the form of $T$ can be found from the second line of~\eqref{GagedPEdeqLinV}.
Consequently, we obtain Proposition~\ref{Prop_PotClass_Quad_G}.

The above consideration implies that  for each admissible transformation of the class~\eqref{PotClass}
the transformation component for~$f$ is of the form~$\tilde f=c_0f$,
where $c_0$ is a~nonzero constant, the transformation component for~$t$ depends only of~$t$
and the transformation component for~$x$ is affine in~$x$.
Therefore, the constraints that single out the subclasses~$\mathcal{P}^1$,~$\mathcal{P}^2$ and~$\mathcal{P}^3$
(i.e. $f_{xxx}\ne0$; $f_{xxx}=0$ and $f\ne\const$; or $f=\const$) are preserved by any admissible transformation of the class~\eqref{PotClass}.
This proves Proposition~\ref{Prop_PotClass_subclasses}.

\section{Comparison of equivalence groupoids}

The equivalence groupoid for the class \eqref{ClassDiffCoef} was presented in~\cite{poch2013d}.
Presenting its description in a~way 
similar to Section~\ref{Section_P_Propositions} we can state the following assertions.

\begin{proposition}\label{Prop_ClassDiffCoef_G}
The usual equivalence group of the class~\eqref{ClassDiffCoef} consists of the transformations
\begin{gather}
\label{ClassDiffCoef_G}
\tilde t=\alpha t+\beta,
\quad
\tilde x=\kappa(x+\mu_1 t+\mu_0),
\quad
\tilde u=\frac{\kappa}{\alpha}(u+\mu_1),
\quad
\tilde f=\frac{\kappa^2}{\alpha}f,
\end{gather}
where $\alpha$, $\beta$, $\kappa$, $\mu_1$, $\mu_0$ are arbitrary constants with $\alpha\ne0$ and $\kappa\ne0$.
\end{proposition}

\begin{proposition}\label{Prop_ClassDiffCoef_NonQuad_G}
The usual equivalence group of the subclass $\mathcal{C}^1=\{\mathcal{C}_f \mid f_{xxx}\ne0\}$ of the class~\eqref{ClassDiffCoef}
coincides with the usual equivalence group of the entire class~\eqref{ClassDiffCoef} and is governed by~\eqref{ClassDiffCoef_G}.
The subclass $\mathcal{C}^1$ is normalized in the usual sense.
\end{proposition}

\begin{proposition}\label{Prop_ClassDiffCoef_Quad_G}
The generalized extended equivalence group of the subclass $\mathcal{C}^2=\{\mathcal{C}_f \mid f_{xxx}=0\}$
of the class~\eqref{ClassDiffCoef} consists of the transformations
\begin{gather*}
\tilde t=\frac{1}{c_0}\!\int\!(X^1)^2\dt+c_5,
\quad
\tilde x=X^1x+\!\int\! (X^1)^2\left(c_1\!\int\!\frac{f^1}{\lambda}\dt+c_3\right)\!\dt+c_4,
\\
\tilde u=c_0\left(\frac{1}{X^1}u-\frac{c_1}{\lambda}x+c_1\!\int\!\frac{f^1}{\lambda}\dt+c_3\!\right),
\quad
\tilde f=c_0f,
\end{gather*}
where $c_0,\ldots,c_4$ are arbitrary constants with $c_0\ne0$ and $(c_1,c_2)\ne(0,0)$,
\begin{gather*}
f=f^2(t)x^2+f^1(t)x+f^0(t),
\quad
X^1(t):=\left(c_1\int \frac{\dt}{\lambda}+c_2\right)^{-1},
\quad
\lambda(t):=e^{2\int f^2\dt}.
\end{gather*}
\end{proposition}

\begin{proposition}\label{Prop_ClassDiffCoef_subclasses}
There are no point transformations between equations from the subclasses $\mathcal{C}^1$ and $\mathcal{C}^2$.
\end{proposition}

Propositions \ref{Prop_ClassDiffCoef_G}--\ref{Prop_ClassDiffCoef_subclasses} give the exhaustive
description of the equivalence groupoid of the class~\eqref{ClassDiffCoef}.
In view of \cite[Section~3.4]{popo2010a} they also imply
that the class \eqref{ClassDiffCoef} has exactly two maximal conditional equivalence groups,
namely the equivalence group of the whole class~\eqref{ClassDiffCoef}
and the generalized extended equivalence group of the subclass $\mathcal{C}^2$.

The equivalence groupoids of the classes~\eqref{PotClass} and~\eqref{ClassDiffCoef} have similar structure.
Propositions \ref{Prop_ClassDiffCoef_G}, \ref{Prop_ClassDiffCoef_NonQuad_G}, \ref{Prop_ClassDiffCoef_Quad_G}
and~\ref{Prop_ClassDiffCoef_subclasses} are the counterparts
of Propositions~\ref{Prop_PotClass_G},~\ref{Prop_PotClass_NonQuad_G},~\ref{Prop_PotClass_Quad_G}
and~\ref{Prop_PotClass_subclasses}, respectively.
This in particular establishes a~correspondence between maximal conditional equivalence groups of these classes,
or, equivalently, equivalence groups (of appropriate kind) of its subclasses,
except the subclass~$\mathcal{P}^3$
of the class~\eqref{PotClass}, which has no counterpart in the class~\eqref{ClassDiffCoef}.

The class~\eqref{GBE} is normalized in the usual sense.
Its usual equivalence group consists of the transformations
\begin{gather*}
\tilde{t}=\frac{\alpha t+\beta}{\gamma t+\delta},
\quad
\tilde{x}=\frac{\kappa x+\mu_1 t +\mu_0}{\gamma t+\delta},
\quad
\tilde{u}=\frac{\kappa(\gamma t+\delta)u-\kappa\gamma x+\mu_1\delta-\mu_0\gamma}{\alpha\delta-\beta\gamma},
\quad
\tilde{f}=\frac{\kappa^2}{\alpha\delta-\beta\gamma}f,
\end{gather*}
where the constant tuple $(\alpha,\beta,\gamma,\delta,\kappa,\mu_1,\mu_0)$
is defined up to a~nonzero multiplier, $\alpha\delta-\beta\gamma\ne0$ and $\kappa\ne0$~\cite{poch2012a}.
Therefore, the class~\eqref{GBE} admits no nontrivial generalizations of the equivalence group
and no nontrivial conditional equivalence groups.

The equivalence group of the class~\eqref{GBE}
is more complicated than the equivalence groups of the classes~\eqref{PotClass} and~\eqref{ClassDiffCoef} since
it additionally contains conformal transformations and hence it is not solvable.
The situation is not the same from the point of view of admissible transformations.
The equivalence groupoid of the class~\eqref{GBE} is of the simplest structure (among all possible ones),
since this class is normalized with respect to its finite-dimensional usual equivalence group,
which is obviously not the case for the classes~\eqref{PotClass} and~\eqref{ClassDiffCoef}.
An explanation for why the class~\eqref{GBE} so differs from the classes \eqref{PotClass} and \eqref{ClassDiffCoef}
in transformational properties is that
an equation of the form~\eqref{GBE} can be potentialized only if $f_{xxx}=0$.

It is interesting to emphasize that the intersection of the classes~\eqref{ClassDiffCoef} and~\eqref{GBE},
which is their subclass consisting of equations of the form $u_t+uu_x+f(t)u_{xx}=0$ with $f\ne0$,
inherits transformational properties of the class~\eqref{GBE}
but not of the class~\eqref{ClassDiffCoef}.
This subclass is also normalized in the usual sense and has the same equivalence group as the whole class~\eqref{GBE}.
The restriction of the generalized extended equivalence group given in Proposition~\ref{Prop_ClassDiffCoef_Quad_G} to the subset
of arbitrary elements~$f$ not depending on~$x$ coincides with the equivalence group for the class~\eqref{GBE}.

\section{Conclusion}

In this paper we have considered the class of generalized potential Burgers equations of the form~\eqref{PotClass}.
Its equivalence groupoid appeared to be of a~complicated structure.
In order to describe this structure in the optimal way, the class~\eqref{PotClass}
has been represented as the disjoint union of three subclasses that are not related by point transformations, namely
\begin{itemize}\setSkips
 \item $\mathcal{P}^1=\{\mathcal{P}_f \mid f_{xxx}\ne0\}$, which is normalized in the usual sense;
 \item $\mathcal{P}^2=\{\mathcal{P}_f \mid f_{xxx}=0,\, f\ne\const\}$, which is normalized in the generalized extended sense;
 \item $\mathcal{P}^3=\{\mathcal{P}_f \mid f=\const\}$, which is normalized in the generalized sense.
\end{itemize}
The partition of the class~\eqref{PotClass} into the three subclasses $\mathcal{P}^1$, $\mathcal{P}^2$
and $\mathcal{P}^3$ results in the partition of its equivalence groupoid into the equivalence groupoids of these subclasses.

The subclass~$\mathcal{P}^3$ is in fact the orbit of the single equation of this subclass with $f=-1$, which is the potential
Burgers equation.
The complexity of the equivalence group of the subclass~$\mathcal{P}^3$ reflects the complexity
of the point symmetry group of a~potential Burgers equation of the form~\eqref{PotClass} with any constant~$f$.
At the same time, the equivalence group of the class~\eqref{PotClass},
which coincides with the usual equivalence group of the subclass~$\mathcal{P}^1$, is rather simple.

Comprehensive analysis has shown that
the equivalence groupoid of the class~\eqref{PotClass} has several  similar features
with the equivalence groupoid of the class~\eqref{ClassDiffCoef}
and completely differs from the equivalence groupoid of the class~\eqref{GBE}.

Any contact admissible transformation between two equations from the class~\eqref{PotClass}
is the first prolongation of a~point transformation between these equations~\cite[Proposition~2]{popo2008d}.
This is why the problem of describing contact transformations in the class~\eqref{PotClass} is reduced to that for point transformations.
Thus, the knowledge of the (point) equivalence groupoid provides us with the exhaustive description of the contact equivalence groupoid.

Knowing transformational properties
of equations from the class~\eqref{PotClass}, which have been collected
in Propositions~\ref{Prop_PotClass_G}--\ref{Prop_PotClass_Quad_G}, is useful
for solving the group classification problem for the class~\eqref{PotClass}.
This problem will be a~subject of a~forthcoming paper~\cite{poch2016a}.

\subsection*{Acknowledgements}

The author thanks Prof. R.O.~Popovych for formulation of the present problem
and collaborating in the course of study a~great number of problems related to generalized Burgers equations.

\itemsep=0ex\parsep=0ex\parskip=0ex\small

\newcommand{\eprint}[1]{#1}

\end{document}